\documentstyle[12pt]{article}

\newcommand{\beq}{\begin{equation}}
\newcommand{\eeq}{\end{equation}}
\newcommand{\eq}[1]{eq. (\ref{#1})}

\title{
Radiative Corrections to the $2E1$ Decay Rate\\
of the $2s$-State in Hydrogen-Like Atoms}

\author{
Savely G. Karshenboim${}^{a,}$\thanks{
E-mail: sgk@onti.vniim.spb.su;
karshenboim@phim.niif.spb.su}
~and Vladimir G. Ivanov${}^b$,\\
\bigskip
{}\\
${}^a$ D. I.  Mendeleev Institute for Metrology,\\
198005, St.~Petersburg, Russia\\
${}^b$ Pulkovo Observatory, 196140, St. Petersburg, Russia
}

\date{}

\begin{document}

\large

\maketitle

\newpage

\begin{abstract}
 Radiative corrections to the $2E1$ decay width of the
 $2s_{1/2}$-state in the low-$Z$ hydrogen-like system are examined
 within logarithmic approximation. The correction is found to be
 $2.025(1)\,\alpha(Z\alpha)^2/\pi\log(Z\alpha)^2$ in units of the
 non-relativistic rate.
\end{abstract}

\newpage

A leading radiative correction to the $E1$ one-photon decay width of the
hydrogenic levels was analyzed recently by the authors in Refs.
\cite{JETP95,IK1}. The correction has order of
$\alpha(Z\alpha)^2\log(Z\alpha)$ in  relative units.  This work is devoted
to the leading radiative correction to the $2E1$ two-photon decay of the
metastable $2s_{1/2}$ state in hydrogen-like atoms, which has the same
relative order of the magnitude.

The rate of the $2s$ metastable state decay in the hydrogen atom was
first examined in Ref.~\cite{Bre} and the investigations were developed
in a number of works (see Refs. \cite{Spi,Kla,Gol,Bac} etc). The decay
of the level in the atom with one Schr\"odinger electron and the infinite
nuclear mass was determined with a high precision in Ref.~\cite{Kla}.
Calculations on the atom with one Dirac electron were performed in
Ref.~\cite{Gol}. The nuclear mass corrections were carried out in
Ref.~\cite{Bac}, where the results of Ref.~\cite{Fri} for the $E1$ decays 
were adjusted for the two-photon $2E1$ transition. However, any radiative
corrections have not yet been discussed previously. In this work a
calculation is made of the leading radiative corrections of the relative
order of $\alpha(Z\alpha)^2$ to the $2E1$ two-photon decay rates in the
low-$Z$ hydrogen-like systems within the logarithmic approximation.

The evaluation of the $\alpha(Z\alpha)^2 \log(Z\alpha)$ correction can be
essentially simplified by using the Yennie gauge \cite{FY}
for virtual photons, in which their propagator in the momentum space has
the form

\[
D^{Y}_{\mu\nu} (k) = \frac{1}{k^2}
\left(
g_{\mu\nu}+2\frac{k_\mu k_\nu}{k^2}
\right).
\]

\noindent
In this gauge it is enough to take into account only the logarithmic
corrections to the energy and the wave functions. Both of them can
be described with using an effective local delta-like potential\footnote{
Relativistic units in which $ \hbar  = c = 1$ and $\alpha=e^2 $ are used
and $Z$ is the nuclear charge in units of the proton one.}

\beq
    V({\bf r}) = \frac{4}{3} \frac{\alpha (Z\alpha)}{m^2}\,
    \log{\frac{1}{(Z\alpha)^2}} \; \delta({\bf r}),
\eeq

\noindent
leading to zero result for all, but $s$ states (see for details
\cite{JETP94,JETP95}).

As it is well known (see e. g. Ref. \cite{IV})
the two-photons decay rate of the $2s$-state in the non-relativistic
approximation is of the form

\beq \label{gtp}
\Gamma_{2\gamma}^{(0)} (2s_{1/2}) = {\cal C} \int\limits_{0}^{1} {dy \,
 y^3(1-y)^3 \left| \sum\limits_{q \neq 2}\, d_{1q}\, d_{2q}\, f_q(y)
 \right| ^2}, \eeq

\noindent
where

\beq \label{const}
{\cal C} =
\frac{4}{3} \frac{\alpha^2}{\pi}
\frac{\Big(E_{2s} - E_{1s}\Big)^5}{(Z\alpha m)^4}
 ,\eeq

\beq \label{dip}
d_{n'n} = \langle n's||Z\alpha m r||np \rangle
,\eeq

\noindent
and the sum has to run over all intermediate $p$-states. For the discrete
$p$-states the function in \eq{gtp} is equal to

\beq      \label{ftp}
f_n(y) = \frac{1}{1-4/n^2+3y} + \frac{1}{4-4/n^2-3y}
.\eeq

The correction can be written as a sum

\beq
\Delta \Gamma_{2\gamma}^{Rad} (2s) = \frac{8}{3} \frac{\alpha
(Z\alpha)^2}{\pi}\, \log{\frac{1}{(Z\alpha)^2}}
\Big(R^{\cal C} + R^{f} + R^{d} \Big)
\Gamma_{2\gamma}^{(0)} (2s)
,\eeq

\noindent
where the contributions labeled by the superscripts ${\cal C}$, $f$  and
$d$ associate with the corrections to the constant ${\cal C}$ (\eq{const}),
the function $f_q(y)$ (\eq{ftp}) and the reduced dipole matrix elements of
\eq{dip}. The results for two first terms presented in Table are easy to
compute in the same techniques as the leading contribution of \eq{gtp}.

The calculation of the third term can be carried out with using
analytic expressions for the radiative corrections to dipole matrix
elements in the Coulomb field. For the discrete $p$-state the expression

\beq              \label{delta_dqn}
 \delta d_{n'n}=
 \frac{\Delta\! E _{1s}}{(Z\alpha)^2 m}
 \frac{(-1)^{n'+1}}{24}
 \sqrt{\frac{n^3(n^2-1)}{n'^3}}
 \times
\]

\[
 \frac{x^4}{\Gamma(n')}
 \sum_{s=0}^{n'}
  y^{n'-s}
  \frac{(-1)^s}{s!}
  \left[ \frac{\Gamma(n'+1)}{\Gamma(n'+1-s)} \right]^2
  \frac{n'-s}{n'}
  \Gamma(n'-s+4)
 \times
\]

\[
  \Biggl\{
   \biggl[
    \psi(n'-s+4)-\psi(n')+2\psi(n'+1)-2\psi(n'+1-s)
    + \log y
\]

\[
- \frac{n'-s}{n+n'} + \frac{4n}{n'(n+n')}
    + \frac{s}{n'(n'-s)} -\frac{3}{2n'}
   \biggr]\, {}_2F_1(n'-s+4,2-n,4,x)
\]

\[
  + {}_2F_{1;a}(n'-s+4,2-n,4,x)
  + \frac{nx}{n'(n+n')}~{}_2F'_1(n'-s+4,2-n,4,x)
  \Biggr\},
\eeq

\noindent
where

\[
 x = \frac{2n'}{n+n'},
\]

\[
y = \frac{2n}{n+n'}~,
\]

\[
 {}_2 F_{1;a} (a_0,b,c,x) =
\left[\frac{\partial}{\partial a} \; {}_2 F_1
(a,b,c,x)\right]_{a=a_0}
,\]

\[
 {}_2 F'_{1} (a,b,c,x) = \frac{\partial}{\partial x}\;
{}_2 F_1 (a,b,c,x),
\]

\noindent and

\[\psi(z) =\Gamma'(z)/\Gamma(z) \]

\noindent
were obtained earlier in Ref.~\cite{IK1}.

One can perform the analytic continuation of the corrections $\delta d_{1n}$
and $\delta d_{2n}$ as functions on $n$ to the
continuous $p$-states.  The sum done over all intermediate $p$-states of
the discrete and continuous parts of the spectrum is evaluated numerically
and the result is summarized in Table.


Our final result of the correction to the decay rate of the
$2s_{1/2}\to 1s_{1/2}$ transition is found to be

\beq
\frac{\Delta\Gamma^{Rad}_{2\gamma}(2s_{1/2})}
{\Gamma^{(0)}_{2\gamma}(2s_{1/2})}
= -2.025(1) \, \frac{\alpha(Z\alpha)^2
}{\pi}  \log{\frac{1}{(Z\alpha)^2}}.
\eeq

\bigskip

\bigskip

This work has been supported in part by grant \# 95-02-03977 of the Russian
Foundation for Basic Research.

\newpage

\begin{center}
\begin{tabular}{||c|c|c||c||}
\hline
\hline
&&&\\[-1ex]
$R^{\cal C}$&$R^{f}$&$R^{d}$&$R^{tot}$\\[1ex]
\hline
\hline
&&&\\[-1ex]
-5.8333 & 2.2115(1) & 2.8625(2) & -0.7593(2) \\[1ex]
\hline
\hline
\end{tabular}
\end{center}

\section*{Caption to Table.}

Radiative corrections to the $2s_{1/2}$ decay rate


\newpage

\begin{thebibliography}{9}
\frenchspacing

\bibitem{JETP95} S. G. Karshenboim, ZhETF {\bf 107} (1995) 1061 /in
Russian/; JETP {\bf 80} (1995) 593.

\bibitem{IK1} V.~G.~Ivanov and S.~G.~Karshenboim,  ZhETF {\bf 109} (1996)
1219 /in Russian/; JETP {\bf 82} (1996)  656; Phys. Lett. {\bf A210} (1996)
313.

\bibitem{Bre} G.~Breit and E.~Teller, Astrophys. J. {\bf 91} (1940) 215.

\bibitem{Spi} L. Spitzer and J. L. Greenstein, Astrophys. J.
{\bf 114} (1951) 407;\\
J. Shapiro and G.~Breit, Phys. Rev. {\bf 113} (1959) 179;\\
B. A. Zon and N. L. Manakov, Pis'ma ZhETF {\bf 7} (1968)
70 /in Russian/; JETP Lett.  {\bf 7} (1968) 52.

\bibitem{Kla} S.  Klarsfeld, Phys. Lett.  {\bf 30A} (1969) 382.

\bibitem{Gol} S. P. Goldman and G.~W.~F.~Drake, Phys. Rev. {\bf A24}
(1981) 183;\\
F.~A.~Parpia and W.~R.~Johnson, Phys. Rev. {\bf A26} (1982) 1142.

\bibitem{Bac} R.~Bacher, Z. Phys. {\bf 315} (1984) 1351.

\bibitem{Fri} Z.~Fried and A.~D.~Martin, Nuovo Cim. {\bf 29} (1963) 574.

\bibitem{FY} H. M. Fried and D. R. Yennie, Phys. Rev. {\bf 112} (1958) 1391.

\bibitem{JETP94} S. G. Karshenboim, ZhETF {\bf 106} (1994) 414 /in
Russian/; JETP {\bf 79} (1994) 230; Yad. Fiz. {\bf 58} (1995)
901 /in Russian/; Phys. At. Nucl. {\bf 58} (1995) 835.

\bibitem{IV} V. B. Berestetskii, E. M. Lifshitz and L. P. Pitaevskii.
{\sl Quantum  Electrodynamics}. Pergamon Press, Oxford (1982).

\end{thebibliography}
\end{document}